\documentclass[aps,prd,twocolumn,superscriptaddress,nofootinbib,reprint]{revtex4-2}
\newcommand\prlsec[1]{\vspace{2mm} \textbf{\emph{#1}}\,---}
\usepackage{amsmath, amssymb}
\usepackage{listings}
\usepackage{graphicx}
\usepackage{dcolumn}
\usepackage{bm}
\usepackage{float,color}
\usepackage{xcolor}
\usepackage[normalem]{ulem}
\usepackage{tabularx}
\usepackage{booktabs}
\usepackage{mathtools} 
\usepackage{multirow}
\usepackage{amssymb}
\usepackage{hyperref}

\begin{document}

\title{
Simultaneous tests of post-Newtonian inspiral using GW250114 and GW250207
}

\author{Saurabh Kumar}\email{saurabhk@cmi.ac.in}
\affiliation{Chennai Mathematical Institute, Siruseri, 603103, India}
\author{Parthapratim Mahapatra}
\affiliation{Department of Physics and Astronomy, School of Physical, Chemical and Environmental Sciences, Cardiff University, Cardiff, CF24 3AA, United Kingdom}
\author{Ish Mohan Gupta}
\affiliation{Department of Physics, University of California, Berkeley, CA 94720, USA}
\author{M. Saleem}
\affiliation{Department of Physics, The University of Texas at Austin, 2515 Speedway, Austin, TX 78712, USA}
\author{B. S. Sathyaprakash}
\affiliation{Institute for Gravitation and the Cosmos, Department of Physics, Penn State University, University Park, Pennsylvania, 16801, USA}
\affiliation{Department of Astronomy and Astrophysics, Penn State University, University Park, Pennsylvania, 16801, USA}
\author{K. G. Arun}
\affiliation{Chennai Mathematical Institute, Siruseri, 603103, India}
\affiliation{Max Planck Institute for Gravitational Physics (Albert Einstein Institute), D-30167 Hannover, Germany}
\affiliation{Leibniz Universit\"at Hannover, D-30167 Hannover, Germany}
\date{\today}

\begin{abstract}
We perform the one of the most general tests of post-Newtonian (PN) inspiral in binary black holes to date by allowing all eight phase coefficients up to 3.5PN order to vary simultaneously from their general-relativistic values. Using two of the loudest binary black hole events observed to date, GW250114 and GW250207, we carry out Bayesian inference with the {\tt IMRPhenomXPHM\_SpinTaylor} waveform model augmented by parametrized deviations in the post-Newtonian phasing. We then identify the combinations of phase coefficients most strongly constrained by the data through a principal component analysis of the eight-dimensional deformation space. We find that the data constrain the three leading principal components beyond their priors, and that all three are consistent with the predictions of general relativity. These results demonstrate that exceptionally loud gravitational-wave events enable robust, high-dimensional tests of inspiral dynamics and provide direct access to the measurable subspace of post-Newtonian deformations.
\end{abstract}
\maketitle

\prlsec{Introduction} GW250114, with a signal-to-noise ratio (SNR) of about 80, and GW250207, with an SNR of about 69, are the two loudest gravitational-wave (GW) events detected to date~\cite{GW250114, GW250114TGR, GW250207}.
These GW events have opened new avenues for testing general relativity (GR) in the strong-field gravity regime~\cite{GW250114,GW250114TGR, Chandra:2025jfc} and provide a glimpse of the capabilities of next-generation GW detectors, which are expected to observe compact binary mergers with SNRs of thousands~\cite{ETScience11,Punturo:2010zz,CE:2019iox,Gupta:2023lga}.  
Therefore, GW250114 and GW250207 provide an excellent testbed to assess the feasibility of gravity tests that have not been possible until now.

One such test is the {\em multiparameter test} of the post-Newtonian approximation to GR (see Ref.~\cite{Bliving2014} for a review), in which one simultaneously searches for departures from GR in multiple PN coefficients that characterize the adiabatic inspiral phase of a compact binary~\cite{AIQS06a,MAIS10,Li:2011cg}. This should be contrasted with the single-parameter tests that look for deviations in individual PN coefficients while assuming that all the remaining coefficients are consistent with the predictions of GR~\cite{BSat94,BSat95,AIQS06a,AIQS06b,YunesPretorius09, MAIS10,Li:2011cg,TIGER,Meidam:2017dgf,FTI}. Since departures from GR may arise at any PN order and potentially in multiple PN orders simultaneously~\citep{Yunes:2016jcc,Endlich:2017tqa,Sennett:2019bpc,Bernard:2022noq}, a multiparameter test, in which all (or a sufficiently large number of) PN deformation parameters are varied together, would be ideal. This class of multiparameter tests is arguably the most generic test of inspiral dynamics, making maximum use of the information encoded in the PN expansion predicted by GR~\cite{YunesPretorius09, GWTC3-TGR}. Moreover, these tests facilitate meaningful mappings between the PN deviation parameters and specific modified theories of gravity~\cite{Bernard:2022noq,Tahura:2018cgk}.

However, performing the most general parametrized PN test, in which all PN deformation parameters are allowed to vary simultaneously, remains challenging within the current phase-only parametrized frameworks, namely Test Infrastructure for General Relativity ({\tt TIGER})~\cite{TIGER,Meidam:2017dgf} and Flexible Theory-Independent ({\tt FTI})~\cite{FTI}. Strong degeneracies among the PN deformation parameters, as well as between these parameters and the intrinsic binary properties, generally lead to poor constraints and broad, uninformative posteriors~\cite{LIGOScientific:2016lio}. This problem could be mitigated either by increasing the sensitivity of the detector network across different frequency bands~\cite{Gupta:2020lxa, Datta:2020vcj} {combining the observations across different frequency bands} or adopting alternative parametrizations based on radiative multipole moments~\citep{Kastha:2018bcr, Kastha:2019brk, Mahapatra:2023hqq, Mahapatra:2023uwd, Mahapatra:2023ydi}.
In the current implementation of parametrized PN tests within the {\tt TIGER} and {\tt FTI} frameworks, principal component analysis (PCA) for the PN deviation parameters has been argued to be a robust alternative, even though the resulting bounds are expressed in terms of a set of linear combinations of the original PN deformation parameters~\cite{AP12,PCA1,PCA3,ShoomEtalSVD,Mahapatra:2025cwk,O4aTGRparam}. 
If the posterior distributions of these parameters indicate a deviation from GR, the relative weights of the individual PN coefficients in these linear combinations can help identify the parameters most likely responsible for the observed deviation. Therefore, as a null test of GR, this provides an efficient method to search for departures from GR in the PN coefficients~\cite{Mahapatra:2025cwk}, and it has now been applied to selected events in the fourth GW Transient Catalog (GWTC-4)~\cite{LIGOScientific:2025slb} released by the LIGO-Virgo-KAGRA (LVK) collaboration~\cite{LIGOScientific:2026qni,O4aTGRparam}. 

\begin{figure*}[t] 
    \centering
    \includegraphics[width=\textwidth]{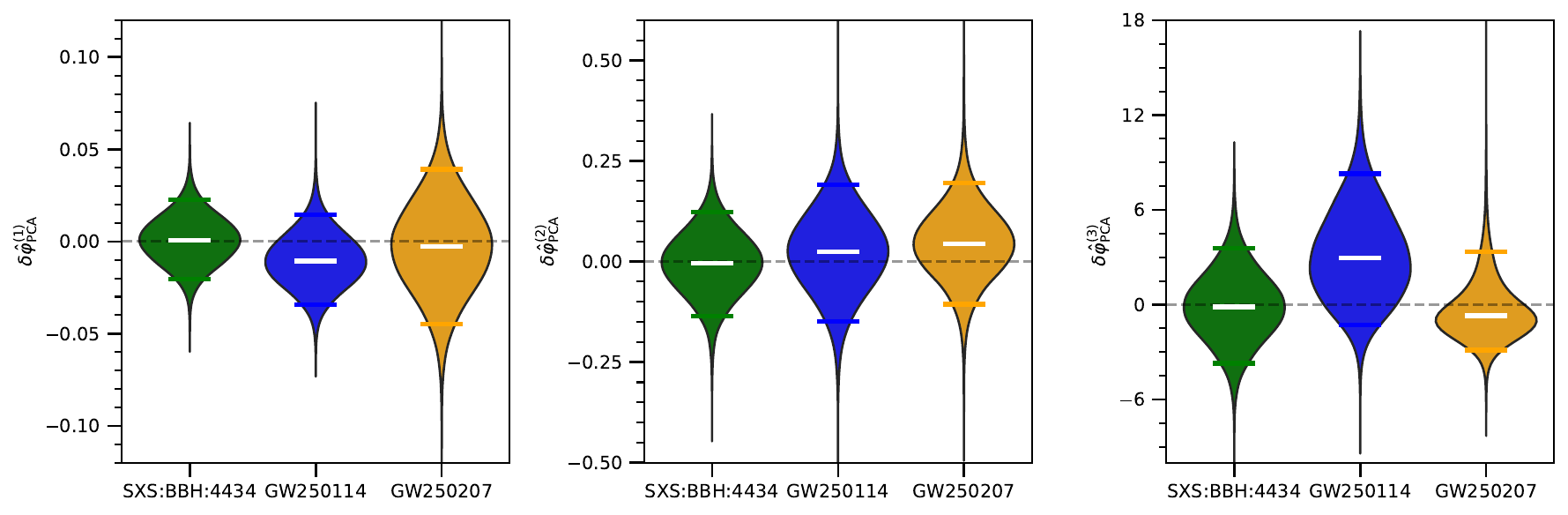}
    \caption{Violin plots of the posteriors of the three PCA parameters 
    obtained from the zero-noise binary black hole (BBH) injection SXS:BBH:4434 waveform (green), GW250114 (blue), and GW250207 (orange). In each violin, the white horizontal bar marks the posterior median and the colored horizontal bars mark the bounds of the 90\% credible interval. The gray dashed line indicates the GR value, $\delta\hat{\varphi}_{\rm PCA}^{(i)}=0$. The results from all three analyses are in good agreement with each other. The composition of the three PCA parameters for different cases can be found in Table~\ref{tab:PCAcomposition}.}
    \label{fig:violin_plot_tiger}
\end{figure*}

Previous works on PCA-based tests of GR~\cite{PCA1,Mahapatra:2025cwk,GW250114TGR,O4aTGRparam} had simultaneously varied fractional PN deformation coefficients from 1.5PN to 3.5PN, resulting in a six-dimensional parameter space whose covariance matrix was diagonalized to construct the principal components.
We excluded the 0PN and 1PN fractional deformation parameters because, for the moderately loud GW signals analyzed so far, they dominate the principal components and substantially reduce sensitivity to deviations entering at higher PN orders~\cite{PCA1,Mahapatra:2025cwk}.
In this paper, we show that GW250114 and GW250207 can facilitate, for the first time, an eight-dimensional test of the adiabatic PN inspiral via PCA, thanks to their high SNRs.\footnote{For GW250207, we use the data with C00 wide calibration, which assumes a wide, uninformative prior for Hanford calibration and the in-situ measured uncertainty envelopes for the other detectors \cite{GW250207}.} The leading three PCA parameters that are precisely inferred carry information about all eight PN coefficients,
providing a more informative and tractable representation of the full eight-dimensional multiparameter test, which is otherwise difficult to interpret in the highly correlated PN parameter space.

\prlsec{Method}
The frequency-domain GW phase of the inspiral waveform from a quasicircular compact binary, obtained using the stationary phase approximation~\cite{SathyaDhurandhar91, DIS00}, for the leading quadrupolar harmonic~\cite{BIOPS2009, ABFO08, MKAF2016}, can be schematically expressed in GR as
\begin{align}\label{eq:phasing}
\Phi_{\rm insp}(f)&=2\pi f\,t_c-\phi_c-\frac{\pi}{4}+\frac{3}{128\,\eta\,v^5} 
\nonumber \\
& \times\quad\left[ \, \sum_{k=0}^{7} \left( \phi_k \, + \phi_{kl} \, \ln v \right) v^k \, + \mathcal{O}(v^{8}) \right] \,.
\end{align}
The coalescence time $t_c$ and coalescence phase $\phi_c$ are kinematical parameters that specify the time and phase of the signal at coalescence. The symmetric mass ratio $\eta$ is defined as the ratio of the reduced mass to the total mass of the binary. In geometrical units (G=c=1), the characteristic orbital velocity parameter is given by $v = (\pi M (1+z) f)^{1/3}$, where $f$ denotes the GW frequency, $M$ denotes the source-frame total mass of the binary, and $z$ denotes the cosmological redshift.
The coefficients $\phi_k$ denote the nonlogarithmic PN phase terms, while
$\phi_{k\ell}$ correspond to the logarithmic contributions that appear
at 2.5PN ($k=5$) and 3PN ($k=6$) order. These coefficients are unique functions of the compact binary's component masses and spins, encoding various physical effects and nonlinearities arising in GR. Although the nonspinning PN phasing for comparable-mass binaries is known up to 4.5PN order~\cite{Blanchet:2023}, current implementations typically introduce phenomenological fractional deformation parameters only up to 3.5PN order~\cite{BDIWW95,BDEI04, GWTC-TGR}, yielding eight independent PN coefficients, including logarithmic contributions at 2.5PN and 3PN order. Deviations from GR are parametrized through fractional deformations of each PN coefficient, $\phi_a \rightarrow (1+\delta\hat{\varphi}_a)\phi_a$, where $a=k,kl$ collectively labels both the logarithmic and nonlogarithmic coefficients appearing in Eq.~(\ref{eq:phasing}),
with all deformation parameters vanishing in GR.

We adopt the {\tt TIGER} framework~\cite{TIGER,Li:2011cg,Roy:2025gzv}, restricting attention to parametrized modifications of the inspiral phase of inspiral--merger--ringdown waveform models. More specifically, we employ a parametrized {\tt  IMRPhenomXPHM\_SpinTaylor} waveform \cite{Colleoni:2024knd}, a frequency-domain inspiral--merger--ringdown model with doubly precessing spins that follows the SpinTaylor prescription for spin precession and includes higher-order spherical harmonic modes of the gravitational radiation~\cite{Pratten:2020ceb,Colleoni:2024knd}. The inspiral phasing is augmented with eight additional parameters that characterize deviations from GR. 
The complete parameter space consists of the two component masses, six spin components, the coalescence time and phase, four angles describing the binary's location and orientation relative to the observer, the luminosity distance, and the eight PN deformation parameters.
 
We used the Bayesian parameter inference package {\tt Bilby TGR}~\citep{bilby_tgr} built upon {\tt Bilby}~\citep{bilby_paper,bilby_pipe_paper}, with the {\tt Dynesty}~\citep{dynesty} sampler (which uses the nested sampling algorithm~\cite{Skilling}) to compute the posterior distribution over this $(15+8)$-dimensional parameter space. 
We employ the prior probability distributions for the 15 source parameters following the standard setup used in the publicly available LVK data~\cite{GWTC3,lvc:dataGWTC3,GWTC3-TGR,LIGOScientific:2025slb,O4aTGRparam}, utilizing priors that are uniform in the component masses with restricted chirp mass ranges of $[20, 40]\,M_\odot$ and $[23, 37]\,M_\odot$ for GW250114 and GW250207, respectively, while applying zero-centered uniform priors bounded between $-30$ and 30 for the fractional PN deformation parameters for both the events. The dimensionless spin magnitudes are taken to be uniform in the range $[0,0.99]$, with isotropically distributed spin orientations.
We then consider the eight-dimensional posterior distribution for the PN deformation parameters, marginalized over all 15 source parameters, and compute its covariance matrix. Diagonalizing this covariance matrix yields the eigenvalues and eigenvectors that define the principal components. The eigenvectors associated with the smallest eigenvalues correspond to the best-constrained combinations of the PN deformation parameters and define the leading informative PCA parameters. The uncertainties in these parameters are determined by the corresponding eigenvalues. To identify the number of PCA parameters that carry significant information from the data, we employ the Jensen--Shannon (JS) divergence~\cite{JSdiv} between the posterior and prior distributions of the PCA parameters. For this, we transform both the posterior samples and samples drawn from the original uniform priors into the PCA basis using the same eigenvector matrix thereby accounting for the nonuniform marginal priors induced by the rotation. We retain only those PCA components for which the JS divergence exceeds 0.1 bits, ensuring that the selected components contain statistically meaningful information gained from the data~\cite{Mahapatra:2025cwk,GW250114TGR,O4aTGRparam}.

The $i^{\rm th}$ PCA parameter has the structure
\begin{equation}
    \delta\hat{\varphi}_{\rm PCA}^{(i)}=\sum_{a}\alpha_{ia} \, \delta\hat\varphi_a,\label{eq:PCA}
\end{equation}
where $\alpha_{ia}$ denotes the weight of the original PN deformation parameter $\delta\hat\varphi_a$ in the $i^{\rm th}$ PCA combination.
We report only the first three ($i=1,2,3$) PCA combinations, for which the JS divergence between the posterior and prior distributions exceeds $0.1$ bits, while higher-order PCA parameters do not contain significant information from the data.
We refer the reader to Ref.~\cite{Mahapatra:2025cwk} for more detailed information on the methodology summarized above.

\prlsec{Results}
Figure~\ref{fig:violin_plot_tiger} summarizes the main results of this paper. The left, middle, and right panels show violin plots of the first three leading PCA parameters for the simulated "zero-noise" numerical relativity (NR) injection and the GW events GW250114 and GW250207, respectively. 
For the NR injection, we adopt the nearly equal-mass nonspinning BBH NR simulation {\tt SXS:BBH:4434}~\cite{Scheel:2025jct}, choosing the total mass, luminosity distance, and other source properties similar to those of GW250114-like system. We inject this NR simulation as a proxy for the true GW signal into the two-detector LIGO network~\cite{ALIGO_det} under the “zero-noise” approximation~\cite{Markovic1993PhRvD} and perform the analysis. As evident from the green curves in Figure~\ref{fig:violin_plot_tiger}, the PCA posteriors sharply peak around zero and recover the GR hypothesis with high confidence. This demonstrates that the employed baseline BBH waveform model {\tt IMRPhenomXPHM\_SpinTaylor} is sufficiently accurate in the nearly equal-mass nonspinning regime, even at such a high SNR ($\sim 80$), and that the subsequent PCA analysis yields unbiased results even for the challenging full eight-dimensional multiparameter test.

In the NR injection, as well as for the GW events GW250114 and GW250207, we find that the JS divergence between the posterior and prior distributions exceeds 0.1 bits for the first three leading PCA parameters.
For both GW events, all the PCA posteriors are statistically consistent with GR.
Beyond their consistency with GR, what is most striking is the remarkable agreement between the NR injection and the two GW events. The composition of the various PCA parameters is reported in Table~\ref{tab:PCAcomposition}, which likewise demonstrates excellent consistency between the NR injection and the two events. 
The median values, 90\% credible intervals, GR quantiles ($\mathcal{Q}_{\rm GR}=P(\delta \hat{\varphi}_{\rm PCA}^{(i)}<0)$), and GR $z$ scores ($z_{\rm GR}$, defined as the number of standard deviations by which the GR value (zero) deviates from the posterior median) for the different PCA posteriors in each case are presented in Table~\ref{tab:PCA90CI_tiger}.
The median values are all consistent with GR, with the uncertainties on the first, second, and third PCA parameters being of ${\cal O}(10^{-2})$, ${\cal O}(10^{-1})$, and ${\cal O}(1)$, respectively. Despite being an eight-parameter test, it is noteworthy that the uncertainty on even the second PCA parameter is of order ${\cal O}(10^{-1})$, highlighting both the power of the PCA method and the amount of information contained in the data.

\begin{table*}[t]
    \centering
    \caption{Composition of the three leading PCA parameters ($\alpha_{ik}$ in Eq.~\ref{eq:PCA}) in terms of the original PN deformation coefficients $\delta\hat\varphi_k$ for the three cases for the 8-parameter (8p) test (varying all of the PN deformation parameters from 0PN to 3.5PN) and for the 6-parameter (6p) test (varying PN deformation parameters from 1.5PN till 3.5PN). }\label{tab:PCAcomposition}
    \renewcommand{\arraystretch}{1.4}
    \resizebox{0.8\textwidth}{!}{
    \begin{tabular}{|c|c|cccccccc|}
    \hline
     & Event & $\delta\hat{\varphi}_0$ & $\delta\hat{\varphi}_2$ & $\delta\hat{\varphi}_3$ & $\delta\hat{\varphi}_4$ & $\delta\hat{\varphi}_{5\ell}$ & $\delta\hat{\varphi}_6$ & $\delta\hat{\varphi}_{6\ell}$ & $\delta\hat{\varphi}_7$ \\
    \hline
    \multirow{2}{*}{$\delta\hat{\varphi}_{\mathrm{PCA}}^{(1)}$} & GW250114 (8p) &  0.8316 &  0.2887 & -0.4360 &  0.0561 & -0.1484 &  0.0921 & -0.0226 & -0.0300 \\
                             & GW250114 (6p) &  - &   - & 0.9107 & -0.1202 & 0.3257 & -0.2069 & 0.0494 & 0.0690    \\
                             & GW250207 (8p) &  0.8404 &  0.2849 & -0.4256 &  0.0539 & -0.1412 &  0.0861 & -0.0215 & -0.0277 \\
                             & GW250207(6p) & - & - &  0.8957 & -0.1235 & 0.3489 & -0.2278 & 0.0529 & 0.0782 \\
                             & SXS:BBH:4434 &  $0.8313$ & $0.2886$ & $-0.4362$ & $0.0562$ & $-0.1491$ & $0.0928$ & $-0.0227$ & $-0.0303$ \\
    \hline
    \multirow{2}{*}{$\delta\hat{\varphi}_{\mathrm{PCA}}^{(2)}$} & GW250114 (8p) &  0.5226 & -0.2184 &  0.6227 & -0.1169 &  0.4051 & -0.3093 &  0.0607 &  0.1192 \\
                             & GW250114 (6p) & - &  - & -0.3967 & -0.0921 & 0.6169 & -0.6097 & 0.0897 & 0.2715 \\
                             & GW250207 (8p) &  0.5105 & -0.2292 &  0.6345 & -0.1171 &  0.4019 & -0.3031 &  0.0601 &  0.1165 \\
                             & GW250207(6p) &  - &  - & -0.4301 & -0.0895 & 0.6089 & -0.5982 & 0.0889 & 0.2656 \\
                             & SXS:BBH:4434 &  $0.5202$ & $-0.2109$ & $0.6154$ & $-0.1176$ & $0.4131$ & $-0.3199$ & $0.0615$ & $0.1250$ \\
    \hline
    \multirow{2}{*}{$\delta\hat{\varphi}_{\mathrm{PCA}}^{(3)}$} & GW250114 (8p) &  0.1770 & -0.4294 &  0.3623 &  0.0360 & -0.4617 &  0.5828 & -0.0561 & -0.3094 \\
                             & GW250114 (6p) & - & -  &  0.1012 & 0.4941 & -0.3417 & -0.2309 & -0.3309 & 0.6827  \\
                             & GW250207 (8p) &  0.1721 & -0.4378 &  0.3495 &  0.0436 & -0.4748 &  0.5782 & -0.0584 & -0.3025 \\
                             & GW250207(6p) &  - &  - & 0.1046 & 0.3785 & -0.4982 & -0.3926 & -0.2203 & 0.6284  \\
                             & SXS:BBH:4434 &  $0.1849$ & $-0.4241$ & $0.3787$ & $0.0347$ & $-0.4516$ & $0.5813$ & $-0.0521$ & $-0.3110$ \\
    \hline
    \end{tabular}}
\end{table*}

Finally, the composition of the leading three PCA components reveals the relative importance of the different PN orders. As expected, the first PCA component is dominated by the 0PN term, followed by the 1.5PN and 1PN terms, and then the 2.5PN logarithmic term.
The second PCA component is dominated by the 1.5PN term, followed by the 0PN term, with subdominant contributions from the 2.5PN logarithmic and 3PN terms, indicating a shift toward higher-order PN contributions to the inspiral phasing. 
The third PCA component is dominated by the 3PN non-logarithmic term, immediately followed by the 2.5PN logarithmic, 1PN and 1.5PN terms, and then the 3.5PN term, further emphasizing the increasing role of higher-order PN terms.
The weakest contributors are the 2PN and 3PN logarithmic terms in the phasing, highlighting the challenges in measuring them in a multiparameter framework. These trends are broadly consistent with Fisher matrix–based projections (see Fig.~2 of Ref.~\cite{PCA2}). The observed hierarchy also depends on the masses and spins of the binary, as discussed in Ref.~\cite{PCA2}.

A comparison of the PCA compositions in the eight-dimensional and six-dimensional (reported in Refs.~\cite{GW250114TGR,GW250207}) multiparameter tests is also presented in Table~\ref{tab:PCAcomposition}. As expected, the inclusion of the first two PN terms in the multiparameter test modifies the dominant contributions to the first two PCA components. However, the overall trend remains unchanged: higher-order PN terms contribute more prominently to the higher PCA components. 
Furthermore, comparing the widths of the PCA posteriors in the two cases, we find that the uncertainties on the first and second PCA parameters are larger by factors of $\sim 2$ and $\sim 4$, respectively, in the six-dimensional test relative to the eight-dimensional one (see Table~\ref{tab:PCA90CI_tiger}). 
However, it should be noted that the hierarchy seen in the composition (see Table~\ref{tab:PCAcomposition}) is specific to the region of parameter space where GW250114 and GW250207 lie and can change for a different class of binaries with different masses, mass ratios, or spins.

One can compare our bounds on the PCA parameters to those from the one-parameter tests performed using TIGER for GW250114 reported in Fig. 5 of \cite{GW250114TGR} and  GW250207 reported in Table 4 of the supplementary material \cite{GW250207}. The best one-parameter bound on $\delta\hat\varphi_0$ for GW250114 (GW250207) is of the order of $4\times10^{-2}$ ($5\times10^{-2}$) which is in good agreement with the bound on our leading PCA parameter, which is roughly $4\times10^{-2}$ ($9\times10^{-2}$). Given the differences in the philosophy of the two classes of tests, further comparisons may not be meaningful.

\begin{table*}[t]
    \centering
    \caption{Constraints on the three leading informative PCA parameters for the GW events GW250114 and GW250207, as well as for the simulated NR injection analysis. We report the median, 90\% credible intervals, the GR quantile $(\mathcal{Q}_{\rm GR})$, and the corresponding $z$-score $(z_{\rm GR})$, for the PCA posteriors.
    }\label{tab:PCA90CI_tiger}
    \renewcommand{\arraystretch}{1.6}
    \resizebox{0.8\textwidth}{!}{
    \begin{tabular}{|c|c|c|c|c|c|c|}
    \hline
    \multirow{2}{*}{Event} & \multicolumn{2}{c|}{properties of  $\delta\hat{\varphi}_{\mathrm{PCA}}^{(1)}$} & \multicolumn{2}{c|}{properties of $\delta\hat{\varphi}_{\mathrm{PCA}}^{(2)}$} & \multicolumn{2}{c|} {properties of $\delta\hat{\varphi}_{\mathrm{PCA}}^{(3)}$} \\
    \cline{2-7}
          & Median \&   & $\mathcal{Q}_{\rm GR}$  & Median \&  & $\mathcal{Q}_{\rm GR}$  & Median \& & $\mathcal{Q}_{\rm GR}$  \\
 
          & $90\%$ errors  & ($z_{\rm GR}$) &  $90\%$ errors   & ($z_{\rm GR}$) &  $90\%$ errors  &  ($z_{\rm GR}$) \\      
    \hline
    GW250114 (8p)   & $-0.01^{+0.02}_{-0.02}$ & 0.76 (0.70)& $0.03^{+0.16}_{-0.18}$ & 0.40 (0.25) & $2.96^{+5.36}_{-4.25}$ & 0.15 (1.00)\\
    GW250114 (6p)   & $0.02^{+0.05}_{-0.06}$ & 0.28 (0.57) & $-0.06^{+0.63}_{-0.64}$ & 0.56 (0.15) & $-3.95^{+17.64}_{-17.56}$ & 0.63 (0.37)\\
    GW250207 (8p)   & $-0.00^{+0.04}_{-0.05}$ & 0.55 (0.12) & $0.04^{+0.16}_{-0.15}$ & 0.31 (0.46) & $-0.68^{+4.06}_{-2.18}$ & 0.66 (0.35)\\
    GW250207 (6p)   & $0.01^{+0.08}_{-0.07}$ & 0.41 (0.23)& $0.07^{+0.60}_{-0.57}$ & 0.42 (0.20)& $7.36^{+13.92}_{-20.63}$ & 0.23 (0.70)\\
    SXS:BBH:4434    & $0.00^{+0.02}_{-0.02}$ & 0.48 (0.06) & $-0.00^{+0.12}_{-0.14}$ & 0.52 (0.05) & $-0.14^{+3.70}_{-3.58}$ & 0.53 (0.06) \\
    \hline
\end{tabular}}
\end{table*}

\prlsec{Conclusions}
In this paper, we have demonstrated that loud events such as GW250114 and GW250207 allow a full-fledged PCA test of GR deviations in the inspiral phasing across the 0PN to 3.5PN orders, including all eight deviation parameters. We have validated this approach through a zero-noise NR injection study and have shown that the leading PCA components from both GW events are consistent with GR. The best-measured PCA combination is constrained at the ${\cal O}(10^{-2})$ level, while remarkably, even the second-best measured combination reaches the ${\cal O}(10^{-1})$ level in this eight-dimensional test. This demonstrates that the PCA-based test provides the most informative alternative currently available to a full multiparameter test, which remains challenging with single-band observations alone.

Our results establish PCA as a powerful framework for testing GR and searching for signatures of beyond-GR physics using loud BBH observations. As the sensitivity of the LIGO, Virgo, and KAGRA detectors~\cite{Acernese_2015,Aasi_2015,KAGRA:2013rdx} improves in future observing runs, PCA-based approaches will enable increasingly stringent tests of adiabatic inspiral dynamics across a broader region of parameter space. This methodology also provides a foundation for exploiting golden GW events in the era of next-generation observatories such as the Einstein Telescope~\cite{ETScience11,Punturo:2010zz} and Cosmic Explorer~\cite{CE:2019iox} to probe gravity in the strong-field regime.

\prlsec{Acknowledgments}
We thank Anuradha Samajdar for an internal review of this document and useful comments on this manuscript. We thank the Testing GR group of the LVK collaboration for several suggestions at different stages of this project.  K.G.A. and S.K. acknowledge support from  Advance Research Grant ANRF/ARG/2025/000931/PS of the Anusandhan National Research Foundation. K.G.A. and S.K. also acknowledge support from the Infosys Foundation. K.G.A. additionally acknowledges support from the Max Planck Society. P.M. acknowledges the Science and Technology Facilities Council for support through Grants ST/V005618/1 and UKRI2489. M.S.\ is supported by the Weinberg Institute for Theoretical Physics at the University of Texas at Austin. The authors are also grateful for computational resources provided by Cardiff University and support by STFC Grants ST/I006285/1, ST/V005618/1 and UKRI2489. The authors are grateful for computational resources provided by the LIGO Laboratory and supported by National Science Foundation Grants No. PHY-0757058 and No.
PHY-0823459.

This research has made use of data obtained from the Gravitational Wave Open Science Center (\url{https://www.gwosc.org/}), a service of the LIGO Laboratory, the LIGO Scientific Collaboration and the Virgo Collaboration. LIGO Laboratory and Advanced LIGO are funded by the United States National Science Foundation (NSF) as well as the Science and Technology Facilities Council (STFC) of the United Kingdom, the Max-Planck-Society (MPS), and the State of Niedersachsen/Germany for support of the construction of Advanced LIGO and the construction and operation of the GEO600 detector. Additional support for Advanced LIGO was provided by the Australian Research Council. Virgo is funded, through the European Gravitational Observatory (EGO), by the French Centre National de Recherche Scientifique (CNRS), the Italian Istituto Nazionale di Fisica Nucleare (INFN) and the Dutch Nikhef, with contributions by institutions from Belgium, Germany, Greece, Hungary, Ireland, Japan, Monaco, Poland, Portugal, Spain. The construction and operation of KAGRA are funded by Ministry of Education, Culture, Sports, Science and Technology (MEXT), and Japan Society for the Promotion of Science (JSPS), National Research Foundation (NRF) and Ministry of Science and ICT (MSIT) in Korea, Academia Sinica (AS) and the Ministry of Science and Technology (MoST) in Taiwan. This material is based upon work supported by NSF’s LIGO Laboratory which is a major facility fully funded by the National Science Foundation. This document has LIGO preprint number LIGO-P2600475.

The data that support the findings of this article are not publicly available. The data are available from the authors upon reasonable request.
\bibliographystyle{apsrev4-2}
\bibliography{ref-list}
\end{document}